\documentstyle[12pt]{article}

\def\dint{\displaystyle\int} 
\def\dsum{\displaystyle\sum} 
 
\newcommand{\beq}{\begin{equation}}
\newcommand{\eeq}{\end{equation}}
\newcommand{\beqa}{\begin{eqnarray}}
\newcommand{\eeqa}{\end{eqnarray}}

\begin{document}
\topmargin 0pt
\oddsidemargin 1mm
\begin{titlepage}
\begin{flushright}
DSFTH 24/98\\
August 1998\\
\end{flushright}
\setcounter{page}{0}
\vspace{15mm}
\begin{center}
{\Large OFF-SHELL AMPLITUDES \\
FOR NONORIENTED CLOSED STRINGS} 
\vspace{20mm}

{\large Luigi Cappiello}, {\large Raffaele Marotta}\\
{\large Roberto Pettorino} and {\large Franco Pezzella~\footnote{e-mail:
Name.Surname@na.infn.it}}\\

{\em Dipartimento di Scienze Fisiche, Universit\`{a} di Napoli\\
and I.N.F.N., Sezione di Napoli,\\
Mostra d'Oltremare, pad. 19, I-80125 Napoli, Italy}\\
\end{center}
\vspace{7mm}

\begin{abstract}
In the context of the bosonic closed string theory, by using the
operatorial formalism, we give a simple
expression of the off-shell amplitude  
with an arbitrary number of external massless states inserted on the 
Klein bottle.
\end{abstract}

\vspace{1cm}

\end{titlepage}
\newpage

In a recent paper \cite{CMPP} we have given a simple prescription for
computing, in the bosonic string theory, the off-shell one-loop amplitude with
an arbitrary number of external massless particles, both for open and
closed oriented strings.

One of the  main reasons for studying off-shell amplitudes is the analysis of
the zero-slope limit ($\alpha^{\prime}\rightarrow 0$) of string theories,
which has to reproduce the perturbative aspects of the ordinary gauge field
theories of fundamental interactions \cite{BK} \cite{DMLRM}. In particular,
closed strings can be used to shed light on perturbative quantum gravity 
\cite{BDS}. Indeed, instead of computing gravity amplitudes by conventional
field theory methods, which are known to be algebraically very complex, one
can use the low-energy limit of the corresponding closed string amplitudes,
which have a very compact expression and which are represented by a single
diagram at each loop.

Nevertheless some {\em stringy} aspects appear when this procedure is
applied to the calculation of amplitudes involving external graviton states.
This is due to the fact that, in the closed string theory, the graviton
belongs to the first excited level together with the dilaton and the
antisymmetric tensor. It is straightforward to disentangle the antisymmetric
tensor from the other two, when regarded as external states, simply by
considering the symmetry property of the polarization tensor. However in the
multiloop amplitudes with external gravitons and dilatons there always remains
the contribution of the antisymmetric tensor states circulating in the loop,
that has to be therefore eliminated if one is interested, as we are,
in  reproducing results in a theory where the antisymmetric tensor is
decoupled from gravitons and dilatons \cite{HV}.

A possibility of solving the above problem is to consider closed strings
propagating also on nonorientable Riemann surfaces which appear in the
perturbative expansion of the nonoriented string. At one-loop level, it
consists {\em to add} the Klein bottle amplitude to the torus one \cite{S}. 
One can show that in so doing the contribution to the amplitude coming from 
the antisymmetric tensor drops out.

This is the main motivation which has prompted us to compute explicitly
off-shell amplitudes on the Klein bottle, extending to it the methods
applied in Ref. \cite{CMPP} to the torus.

In this letter we construct amplitudes for massless states of the closed bosonic
string on the Klein bottle by using the operatorial formalism of 
the $N$-string $g$-loop Vertex \cite{DPFLHS}
which provides, as shown in \cite{CMPP}, a natural prescription leading to an
off-shell extension of amplitudes.

Our starting point is the well-known fact that it is possible to
rewrite on-shell closed string amplitudes on the Klein bottle in terms 
of on-shell open string amplitudes \cite{S} \cite{HKKS}. 
This suggests to use
the $N$-string $g$-loop Vertex for open strings \cite{DPFLHS} in 
the one-loop case,
i.e. $g=1$, in order to
obtain the relative off-shell extension. After some manipulations,
we arrive at an expression for off-shell amplitudes involving massless states
which have the same structure as the corresponding ones on the torus given in 
\cite{CMPP}; moreover it naturally involves geometrical quantities related to 
the Klein bottle.

The nonoriented closed string theory is obtained by requiring invariance under
the world-sheet parity symmetry. In the Hilbert space of the physical
states, this corresponds to the action of the ``turnover'' operator \cite{S}
\cite{HKKS}:

\begin{equation}  \label{TUR}
T:\;\alpha_m^\mu \longleftrightarrow \bar{\alpha }_m^\mu  .
\end{equation}

States which are odd under the action of $T$ drop out from the spectrum and
are annihilated by the projection operator:
 
\[
P_{+}=\displaystyle\frac{1}{2}(1+T) .
\]

One-loop amplitudes for nonoriented strings are obtained by inserting the
projection operator $P_{+}$ into the trace over the internal states. In this
case the first term of the projector obviously gives the torus amplitude
considered in \cite{CMPP}, while the second one gives the Klein bottle
contribution to the amplitude, which we are going now to analyze. When the
two terms are added together, as implied by the insertion of $P_{+}$, the
internal states which are odd under the action of $T$ are cancelled out and
the only contribution comes from states which are symmetric under the $%
\alpha \longleftrightarrow \bar{\alpha}$ exchange. In particular, among the
massless states, only the dilaton and graviton will circulate in the loop.
It is evident that if one subtracted the Klein bottle amplitude from the
torus one by inserting 
\[
P_{-}=\displaystyle\frac{1}{2}(1-T),
\] 
only antisymmetric states would contribute.
 
The Klein bottle on-shell amplitude for $M$ arbitrary external states 
is given by:

\begin{equation}
A_{KB}^{(1)}(p_{1}\,\hspace{-0.01in},\cdots ,p_{M}) \sim
\int d^{d}p\ {\mbox Tr}_{\alpha\bar{\alpha}}
\left\{ 
\displaystyle\prod\limits_{i=1}^{M} \left[ \displaystyle \int\limits_{|x_{i}|%
\leq 1}{}\displaystyle\frac{d^{2}x_{i}}{|x_{i}|^{4}} x_{i}^{L_{0}}
\bar{x}_{i}^{\bar{L}_{0}} W_{\alpha \bar{\alpha}}(i)  
\right] T \right\}
\label{KBON} 
\end{equation}
for a particular cyclic ordering. 

In the equation (\ref{KBON})
\[
W_{\alpha \bar{\alpha}}(i) = {\cal N}_{i} :V_{\alpha }(\frac{p_{i}}{2}\,,\,x)
\bar{V}_{\bar{\alpha}}(
\frac{p_{i}}{2}\,,\bar{x}): 
\]
represents the closed string vertex operator for the $i$-th 
on-shell closed string state,
factorized in two open string vertices, $V_{\alpha}$ and
$\bar{V}_{\bar{\alpha}}$, each carrying
half the momentum of the state
and $L_{0}$ is the usual Virasoro generator including both the orbital
and the ghost contribution. The normalization constant ${\cal N}_{i}$
is peculiar of the external state in consideration.

The vertex operator associated to a massless state is given by:
\beqa
W(p;x,\bar{x}) & = &i {\cal N}_G \epsilon_{\mu \nu}:\partial_x X^\mu
(x)\partial _{\bar{x}}\bar{X}^\nu (\bar{x})e^{i\sqrt{2\alpha
^{\prime }}p\cdot X(x,\bar{x})}: \nonumber \\
& = & i {\cal N}_{G} :\xi_{\mu} \partial_x X^\mu 
e^{i\sqrt{2\alpha^{\prime}} \frac{p}{2} \cdot X} \bar{\xi}_{\nu}
\partial_{\bar{x}} \bar{X}^\nu e^{i\sqrt{2\alpha^{\prime}}
\frac{p}{2} \cdot \bar{X}}:          \label{VER}
\eeqa
where the polarization tensor, due to the structure of the vertex operator,
has been considered as:
\[
\epsilon_{\mu \nu} = \xi_{\mu} \otimes \xi_{\nu} .
\]
It defines an on-shell massless state, satisfying the conditions:
\beqa
p^2=0 &{~~~~~~~~~~}&\epsilon \cdot
  p =0  \label{onshell}   ,
\eeqa
through:
\beqa
|\epsilon; p> & = &\lim _{x,\bar{x}\rightarrow 0}W(p;x, 
\bar{x})|0>   \nonumber \\
& = & {\cal N}_{G} \epsilon_{\mu \nu} \alpha^{\mu}_{-1} \bar{\alpha}^{\nu}_{-1}
|p>  .\label{O-S}
\eeqa

The effect of $T$ in (\ref{KBON}) consists in transforming the trace 
over $\alpha $ and $\bar{%
\alpha}$ into the trace over a single set of oscillators, as in the case of
an open string amplitude \cite{S} \cite{HKKS}, i.e.: 
\begin{equation}
A_{KB}^{(1)}(p_{1},\cdots, p_{M})\sim\,
\displaystyle{}\prod\limits_{i=1}^{M} \left[ {\cal N}_{i}\! \! \!\ \! \!\!\!\!
\int
\limits_{|x_{i}|\leq 1}{}\displaystyle\frac{d^{2}x_{i}}{|x_{i}|^{4}}\right] %
\displaystyle\int d^{d}p {\mbox Tr}_{{}\alpha }\left\{ \displaystyle%
\prod\limits_{i=1}^{M}\left( x_{i}^{L_{0}}V_{\alpha}(\frac{p_{i}}{2}%
,{}1)\right) \displaystyle\prod\limits_{i=1}^{M}\left( \bar{x}_{i}^{L_{0}}
V_{\alpha }(\frac{p_{i}}{2},{}1)\right) \right\}.
\label{AAKK}
\end{equation}
Some straightforward calculations allow us to rewrite the amplitude in
eq. (\ref{AAKK}) in the
form:
\[
A_{KB}^{(1)}(p_{1},\cdots, p_{M})=C_{KB}^{(1)} \left[ \prod_{i=1}^{M}
{\cal N}_{i} \right]
\displaystyle\int \displaystyle\frac{d^{2}k}{|k|^{4}}k^{M}\displaystyle%
\prod\limits_{i=1}^{M-1}\left[ \displaystyle\int\limits_{|k|\leq |z_{i}|\leq
1}d^{2}z_{i}\right] 
\displaystyle\int d^{d}p 
\]
\begin{equation}
\times   
{\mbox  Tr}_{\alpha }\left\{ 
V_{\alpha }\left( \displaystyle
\frac{p_{M}}{2}\,,1 \,\right)V_{\alpha }\left( \displaystyle\frac{p_{1}}{2}%
\,,\,z_{1} \right) \cdots 
 V_{\alpha }\left( \displaystyle
\frac{p_{M-1}}{2}\,,z_{M-1} \,\right)
\right.
\label{KB3}
\end{equation}
\[
\left.
 V_{\alpha }\left( \displaystyle\frac{p_{M}}{2}%
\,,\,k\right) V_{\alpha }\left( \displaystyle\frac{p_{1}}{2}%
\,,\,k \bar{z}_{1} \right) \cdots V_{\alpha }\left( \displaystyle
\frac{p_{M-1}}{2}\,,k\bar{z}_{M-1} \,\right) |k|^{2L_{0}}\right\}
\]
where 
\[
z_{r} = x_{1} \cdots x_{r}
\]
\[
k \equiv z_{M}  = x_{1} \cdots x_{M} .
\]
The normalization constant $C_{KB}^{(1)}$ of the amplitude,
which depends on the topology of the Riemann surface, will be determined in
the following in terms of the torus one. 
 
Eq. (\ref{KB3}) shows that the problem of computing an
amplitude involving $M$ external states 
on the Klein bottle is reduced to the evaluation of a planar amplitude
involving $2M$ open string states at the one-loop order, provided 
that $k$ is substituted in the
integrand by $|k|^2$ and that the first $M$ vertices, carrying the 
momentum $p_i$ (and with polarization $\xi_{i}$ in the case
of massless states) are inserted
at the complex punctures $z_i$ while the
$M$ remaining ones, still with momentum $p_i$ (and with polarization
$\bar{\xi}_i$), are inserted at the image points 
$k\bar z_i$ .

It then follows that we can use the $N$-string $g$-loop
Vertex $V_{N;g}$ \cite{DPFLHS} of the open string theory to 
reproduce the on-shell amplitude (\ref{KB3}) for 
external massless states, specializing
it to the case $N=2M$ and $g=1$ and, more importantly, to get an off-shell
extension, following the procedure of Ref. \cite{CMPP}.

The Vertex $V_{N;g}$ depends on $N$ complex Koba-Nielsen variables 
$z_{i}$'s,
corresponding to the punctures of the external states, through $N$
conformal transformations $V_{i}$'s, which define a local coordinate
system vanishing around each $z_{i}$, i.e.:
\[
V_{i}(0)=z_{i}.
\]
When $V_{N;g}$ is saturated with $N$ physical string states satisfying the
mass-shell condition, the corresponding amplitude does not depend on the
$V_{i}$'s. If this condition is relaxed, the dependence of
$V_{N;g}$ on them is transferred
to the off-shell amplitude. This is analogous to what happens
in gauge theories, where on-shell amplitudes are gauge invariant, while
their off-shell counterparts are not.

At the one-loop order a consistent choice for $V_{i}(z)$ was shown
to be \cite{CMPP} 
\[
V_{i}(z)=z_{i}e^{z}.
\]
This choice leads, in the open string case, 
to the following $N$-string one-loop Vertex $V_{N;1}$ ready to be saturated with
massless states and that we
are going to use for computing our amplitudes: 
\[
V_{N;1}\, =\, {\cal C}_{1}^{open}\! < \Omega| \int [ {\mbox d} m]^{1}_{N}
\]
\begin{equation}
\times \exp \bigg\{ \sum_{\stackrel{i,j=1}{i \neq j}}^{N} 
\left[ \sqrt{2 \alpha'} p^{(i)} + \alpha_{1}^{(i)} z_{i}
\partial_{z_{i}} \right] \cdot
\left[ \sqrt{2 \alpha'} p^{(j)} + \alpha_{1}^{(j)}
z_{j} \partial_{z_{j}} \right]  G_{o}(z_{i},z_{j} ) \bigg\} 
\label{VMP1}
\end{equation}
where $< \Omega|$ denotes the bra-vacuum and $G_{o}(z_{i},z_{j})$ is the
right translational invariant open string Green 
function, given by \cite{M,RS}:
\[
G_{o}(z_{i},z_{j})={\cal G}_{o}(z_{i},z_{j})-\displaystyle\frac{1}{2}\log
V_{i}^{\prime }(0)-\displaystyle\frac{1}{2}\log V_{j}^{\prime }(0)
\]
with \cite{DPFLHS}:
\begin{equation}  \label{OGR}
{\cal G}_o(z_i,z_j)=\ln \left[ (z_i-z_j)\displaystyle\prod\limits_{n=1}^{+\infty } 
\frac{(z_i-z_j|k|^{2n})(z_j-z_i|k|^{2n})}{z_i\,z_j(1-|k|^{2n})^2}\right] + %
\displaystyle\frac{\log {}^2\left( \displaystyle\frac{z_i}{z_j}\right) }{%
4\log |k|}
\end{equation}   
\vspace{.5cm}
By adapting $V_{N;1}$ to the Klein bottle case with $2M$ massless states
one gets:
\[
A_{KB}^{(1)}(p_{1},\cdots, p_{M})=({\cal N}_{G})^{M}C_{KB}^{(1)}\displaystyle\int
\left[ dm_{KB} \right] _{M}^{(1)} k^M 
\displaystyle\prod\limits_{i\neq j=1}^{2M}\
\exp \left\{ \displaystyle\frac{\alpha ^{\prime }}{2}p_{i}\cdot
p_{j}G_{o}(z_{i},z_{j})\right\}
\]
\begin{equation} 
\times \exp \left\{ \displaystyle 2\sum_{\stackrel{i,j=1}{i \neq j}}^
{2M} \sqrt{
\displaystyle\frac{\alpha ^{\prime }}{2}}\xi _{i}\cdot p_{j}\partial
_{z_{i}}G_{o}(z_{i},z_{j})+\displaystyle\sum_{\stackrel{i,j=1}{i \neq j}}^{2M}
\xi
_{i}\cdot \xi _{j}\partial _{z_{i}}\partial
_{z_{j}}G_{o}(z_{i},z_{j})\right\}
\label{OKB}
\end{equation}
where $z_{M+i}=k \bar{z}_{i}$, $\xi _{M+i}=\bar{\xi}_{i}$, 
$p_{M+i}=p_{i}$ for $i=1,\cdots, M$ and $z_M=1$ \cite{RS}.  

The eq. (\ref{OKB}) has to be understood as an expansion in $\xi _{i}$
restricted to the terms which are linear in any $\xi _{i}$.

The measure turns out to be:
\beq
\left[ dm_{KB}\right] _{M}^{(1)}=\prod\limits_{i=2}^{M}d^{2}z_{i}\,%
\displaystyle\frac{d^{2}k}{|k|^{4}}\left[ -\ln |k|\right] ^{-d/2\,}%
\displaystyle\prod\limits_{n=1}^{+\infty }\left( 1-|k|^{2n}\right) ^{2-d} .
            \label{meas}
\eeq

The constant $C_{KB}^{(1)}$ can be determined by looking at the partition
functions on the torus and on the Klein bottle, which in turn can be
straightforwardly determined from the expressions of the measures in the two
cases:

\[
\begin{array}{c}
Z_T^{(1)}=C_T^{(1)}\displaystyle\int \displaystyle\frac{d^2k}{|k|^4}\left[
-\ln |k|\right] ^{-d/2\,}\displaystyle\prod\limits_{n=1}^{+\infty }\left(
\left| 1-k^n\right| ^2\right) ^{2-d} \\ 
=C_T^{(1)}\displaystyle\int \displaystyle\frac{d^2k}{|k|^4}\left[ -\ln
|k|\right] ^{-d/2\,}\left[ 1 - k - \bar{k} + (d-2)^2|k|^2+O(|k|^{3}) \right] ,
\end{array}
\]

\[
\begin{array}{c}
Z_{KB}^{(1)}=C_{KB}^{(1)}\displaystyle\int \displaystyle\frac{d^2k}{|k|^4}%
\left[ -\ln |k|\right] ^{-d/2\,}\displaystyle\prod\limits_{n=1}^{+\infty
}\left( 1-|k|^{2n}\right) ^{2-d} \\ 
=C_{KB}^{(1)}\displaystyle\int \displaystyle\frac{d^2k}{|k|^4}\left[ -\ln
|k|\right] ^{-d/2\,}\left[ 1+(d-2)|k|^2+ O(|k|^{4}) \right] .
\end{array}
\]

The coefficient of the linear term in $|k|^2$ counts the physical
massless states.
If one requires the combination 
$\displaystyle\frac{1}{2}(Z_T^{(1)}+Z_{KB}^{(1)})$ to give the right 
counting $\displaystyle\frac{(d-2)(d-1)}2$ of the states circulating in the loop, 
i.e. the symmetric ones,
one has to
take: 

\[
C_{KB}^{(1)}=C_T^{(1)} .
\]

The expression (\ref{OKB}), although already in a quite compact form, does
not make explicit the relation with the geometric properties of the Klein
bottle. However, we show that it is possible to rewrite it in
terms of geometrical quantities defined on the Klein bottle. After some
algebraic calculations, indeed one gets:
\vspace{.5cm}
\[
A_{KB}^{(1)}(p_{1},\cdots,p_{M})=  ({\cal N}_{G})^{M}C_{KB}^{(1)}\displaystyle\int
\left[ dm_{KB}\right] _{M}^{(1)} \times \exp \left\{ -2\displaystyle%
\sum\limits_{i=1}^{M}\xi _{i}\cdot \bar{\xi}_{i}\partial _{z}\partial
_{\bar{z}} G_{KB}(z,z_{i})|_{z=z_{i}}\right\} 
\]
\begin{equation}  
\times \exp \left\{ \sum_{\stackrel{i,j=1}{i \neq j}}^{M}\left[ \sqrt{\frac{\alpha
^{\prime }}{2}}p_{i}+\xi _{i}\partial _{z_{i}}+\bar{\xi}_{i}\partial _{\bar{z%
}_{i}}\right] \cdot \left[ \sqrt{\frac{\alpha ^{\prime }}{2}}p_{j}+\xi
_{j}\partial _{z_{j}}+\bar{\xi}_{j}\partial _{\bar{z}_{j}}\right]
G_{KB}(z_{i},z_{j})\right\} 
\label{KBA}
\end{equation}
where $G_{KB}(z_{i},z_{j})$ can be given in terms of either the open string
Green function $G_{o}$ or the closed string one $G_{c}$
at the one-loop order \cite{GSW} as follows:

\begin{eqnarray}
G_{KB}(z_{i},z_{j})& = & G_{o}(z_{i},z_{j})+G_{o}(z_{i},k\bar{z}_{j})+G_{o}(k%
\bar{z}_{i},z_{j})+G_{o}(k\bar{z}_{i},k\bar{z}_{j}){}  \label{COF} \\
& & -G_{o}(z_{i},k\bar{z}_{i})-G_{o}(z_{j},k\bar{z}_{j})  \nonumber \\
&=&G_{c}(z_{i},z_{j})+G_{c}(k\bar{z}_{i},z_{j})+G_{c}(z_{i},k\bar{z}_{j})
+G_{c}(k\bar{z}%
_{i},k\bar{z}_{j})  \nonumber \\
&&-G_{c}(z_{i},k\bar{z}_{i})-G_{c}(z_{j},k\bar{z}_{j}) .  \label{CGF}
\end{eqnarray}

The Green function $G_{c}(z_{i},z_{j})$ is expressed in terms of the
real modular parameter $|k|^{2}$:
\begin{equation}
G_{c}(z_{i},z_{j})=\frac{1}{2}\log \left| \frac{(z_{i}-z_{j})}{\sqrt{%
z_{i}\,{}\,z_{j}}}\prod\limits_{n=1}^{\infty }\frac{\left( z_{i}-\left|
k\right| ^{2n}z_{j}\right) \left( z_{j}-\left| k\right| ^{2n}z_{i}\right) }{%
z_{i}z_{j}\left( 1-\left| k\right| ^{2n}\right) ^{2}}\right| ^{2}+\frac{1}{4}%
\frac{\log ^{2}\left| z_{i}/z_{j}\right| }{\log \left| k\right| }
\end{equation}

Notice that $G_{KB}(z_{i},z_{j})$ is invariant under the transformation $%
z_{i}\rightarrow k\bar{z}_{i}$, which is reminiscent of the involution
which is used to define the Klein bottle in terms of the double
covering torus $\cite{BM}$. 
 However, to make a complete correspondence with
this geometrical representation one has to take into account that the 
Klein bottle is described by a double covering torus with a real
modular parameter. In our formalism this parameter can be identified 
with the modulus of $k$, while the phase of $k$ will be reinterpreted 
as the coordinate of one of the punctures. 
This can be easily accomplished by making the
change of variable $z_{i}\rightarrow e^{i\pi \tau _{1}}z_{i}$ , where we have
parametrized $k=e^{2i\pi (\tau _{1}+i\tau _{2})}$ as usual. As the Green
function $G_{KB}(z_{i},z_{j})$ only depends on the ratio of its arguments, it is
easy to see that after the substitution, it will depend only on the modulus
of $k$. 
 Moreover, after that change of variable, the Green 
function $G_{KB}(1,z_{j})$, which is also present 
in eq. (\ref{KBA}), becomes equal to $G_{KB}(e^{-i\pi \tau_1},z_{j})$. This 
allows us to reinterpret the first argument, $e^{-i\pi \tau_1}$, as the 
position of one of the punctures, which we call again $z_1$, with
modulus fixed to one (a similar 
discussion can be found in the first two references in \cite{HKKS}).     
It is also clear that the integration measure and the integration
region over the punctures remain unchanged. 

After these manipulations the expression of the amplitude has still the form
given in the eq. (\ref{OKB}), but with the following Green function and measure:
\begin{eqnarray}
G_{KB}(z_{i},z_{j}) &=&G_{0}(z_{i},z_{j})+G_{0}(z_{i},\left| k\right| \bar{z}%
_{j})+G_{0}(\bar{z}_{j},\left| k\right| z_{i})+G_{0}(\bar{z}_{i},\bar{z}_{j})
\nonumber \\
{} &&-G_{0}(z_{i},\left| k\right| \bar{z}_{i})-G_{0}(z_{j},\left| k\right| 
\bar{z}_{j})  \nonumber \\
{} &=&\displaystyle G_{c}(z_{i},z_{j})+G_{c}(\bar{z}_{j},\left| k\right|
z_{i})+G_{c}(z_{i},\left| k\right| \bar{z}_{j})+G_{c}(\bar{z}_{i},\bar{z}_{j}) 
\nonumber \\
&&-G_{c}(z_{i},\left| k\right| \bar{z}_{i})-G_{c}(z_{j},\left| k\right| \bar{z}_{j})
\label{KBG}
\end{eqnarray}
\[
\left[ dm_{KB}\right] _{M}^{(1)}=2i \frac{d^{{}}\left| k\right| }{|k|^{3}}%
\frac{dz_1}{z_1}\prod\limits_{i=2}^{M}d^{2}z_{i}\,\displaystyle\left[ -\ln |k|\right]
^{-d/2\,}\displaystyle\prod\limits_{n=1}^{+\infty }\left( 1-|k|^{2n}\right)
^{2-d}  
\]

Our expression for the Klein bottle Green function is still different from
the one given in Ref. \cite{BM}, where the authors use a sort of image
method to compute it as a combination of Green
functions defined on the double covering torus. In fact the 
difference consists in the presence in (\ref
{KBG}) of the last two terms which contain the torus
Green function evaluated at conjugate points. These two terms are by
themselves invariant under the involutions $z\longrightarrow \left| k\right| 
\bar{z}_{i}$ and appear as a sort of ``normal ordering'' at coincident
points on the Klein bottle.

The eq. (\ref{KBA}) giving the $M$-point amplitude of massless states
on the Klein bottle is formally identical to the corresponding one on
the torus \cite{CMPP}, modulo the replacements of the Green function on the
torus with $G_{KB}(z_{i},z_{j})$  and the torus measure with $\left[
dm_{KB}\right] _{M}^{(1)}$ in (\ref{meas}).

Let us now specialize to the two-point amplitude, which is obtained by
expanding, as in the case of the torus, the eq. (\ref{KBA}) in terms
linear in the polarizations $\epsilon _1^{\mu \nu }=$$\xi _1^\mu \otimes 
\bar{\xi }_1^\nu $ and $\epsilon _2^{\mu \nu }=$$\xi _2^\mu \otimes \bar{
\xi }_2^\nu $. After some straightforward algebra completely analogous to the
one done in that case in Ref. \cite{CMPP} one arrives at the following
expression for the two-point one-loop amplitudes on the Klein bottle: 

\beqa
T_{\mu \nu \rho \sigma} &  =  & \bigg\{ - \frac{2}{p^{2}} (a_{3} + a_{2} )
\left[ \eta_{\mu \rho} p_{\nu} p_{\sigma} + \eta_{\nu \rho} p_{\mu} p_{\sigma}
+ \eta_{\mu \sigma} p_{\rho} p_{\nu} + \eta_{\nu \sigma} p_{\rho} p_{\mu}
\right]  \nonumber \\
& {} & - \frac{4}{p^{2}} a_{1} \left[ \eta_{\mu \nu} p_{\rho} p_{\sigma} +
\eta_{\rho \sigma} p_{\mu} p_{\nu} \right] + \frac{4}{p^{4}} (a_{1}+a_{2}
+a_{3}) p_{\mu} p_{\nu} p_{\rho} p_{\sigma} \nonumber \\
& {} & +  2 (a_{3}+a_{2}) \left[ \eta_{\mu \rho} \eta_{\nu \sigma} +
\eta_{\nu \rho} \eta_{\mu \sigma} \right] + 4 a_{1} \eta_{\mu \nu}
\eta_{\rho \sigma} \bigg\} \nonumber \\
& {} & + \bigg\{ - \frac{2}{p^{2}} (a_{3} - a_{2} ) \left[ \eta_{\mu \rho}
p_{\nu} p_{\sigma} - \eta_{\nu \rho} p_{\mu}p_{\sigma} + \eta_{\nu \sigma}
p_{\mu} p_{\rho} - \eta_{\mu \sigma} p_{\rho} p_{\nu} \right] \nonumber \\
& {} & + 2 (a_{3} - a_{2} )  \left[ \eta_{\mu \rho} \eta_{\nu \sigma}
- \eta_{\mu \sigma} \eta_{\nu \rho} \right] \bigg\} \ \equiv \ S_{\mu \nu
\rho \sigma} + A_{\mu \nu \rho \sigma} 
\eeqa
where (we use the same notation as in \cite{CMPP}):
$$
\begin{array}{ll}
a_1=\dint \left[ dm_{KB} \right]_{2}^{1} e^{{\alpha}^{\prime} p^{(1)} \cdot p^{(2)}
G_{KB}(z_{1},z_{2})} \partial_{z_{1}} \partial_{\bar{z}_{1}}
G_{KB}(z_{1},z_{2}) \partial_{z_{2}} \partial_{\bar{z}_{2}} G_{KB}(z_{1},z_{2}) \\

a_2=\dint \left[ dm_{KB} \right]_2^{1} e^{\alpha ^{\prime} p^{(1)} \cdot p^{(2)}
G_{KB}(z_{1},z_{2}) } \partial_{z_{1}} \partial_{\bar{z}_{2}}
G_{KB}(z_{1},z_{2}) \partial_{\bar{z}_{1}} \partial_{{z}_{2}} G_{KB}(z_{1},z_{2}) \\

a_3=\dint \left[ dm_{KB} \right]_2^{1} e^{\alpha ^{\prime} p^{(1)} \cdot p^{(2)}
G_{KB}(z_{1},z_{2}) } \partial_{z_{1}} \partial_{{z}_{2}}
G_{KB}(z_{1},z_{2}) \partial_{\bar{z}_{1}} \partial_{\bar{z}_{2}} G_{KB}(z_{1},z_{2}) 
\end{array}
$$
and where we have considered the momentum conservation for setting $p^{(1)}=-p^{(2)} \equiv p$
and have made explicit the symmetry properties on the indices $(\mu \nu)$ which
refer to the polarization tensor of the particle (1)  and on 
$(\rho \sigma)$ which refer to the one of the particle (2). Of course
the amplitude is symmetric under the exchange of the states $(1)$ and
$(2)$. 

It is remarkable that this amplitude, as an off-shell string amplitude, is 
{\em transverse},
 i.e.:
\[
p^{\mu} T_{\mu \nu \rho \sigma} = 0  .
\]

The same property was shown to hold for the torus amplitudes too in \cite{CMPP}.

It is well-known that in quantum gravity the transversality of the
amplitudes
follows when background-field techniques are used, then our off-shell 
amplitudes, when added to the torus ones, should 
give in the field theory limit $(\alpha'\rightarrow 0)$
one-loop corrections to quantum gravity coupled to scalars,
obtained in the framework of the background field method.

Finally we would like to make a remark on a possible reinterpretation of the 
procedure we have used in this paper to construct one-loop Klein bottle 
amplitudes
and make a connection with the so-called sewing technique which can be used in the framework of the
operatorial formalism to construct multiloop string amplitudes.

As we have repeatedly stressed, the final form of the amplitude is nicely
written in terms of objects
related to the geometry of the Klein bottle surface. Since, however, our
starting point was the use of the 
operator formalism, and in particular the insertion of the $turnover$
operator $T$ into the trace on the internal states 
circulating in the loop, it would be desirable to establish from the
beginning a connection between this approach 
and the geometry of the Klein bottle.

Let us first remark that $T$ is defined in eq. (1) as acting on the
oscillator operators. On a Riemann surface these 
operators appear in the $local$ expansion of the string field around a given
puncture $z_i$ provided with  sets of 
complex coordinates $\zeta _i=V_i^{-1}(z)$, which map a neighborhood of the
origin in the complex plane into a  neighborhood of 
$z_i$ . On the other hand, it is well-known that any genus $g$
Riemann  surface with $N$ punctures can be 
obtained from an $(N+2g)$-punctured sphere in which $2g$ punctures  are
suitably 
identified through the so-called $sewing$ procedure, so that $g$ handles
are generated. 
In the following we will consider for simplicity only the one-loop
case, $g=1$. 

To identify the two 
punctures $z_i$ and $z_j$ means to make a correspondence 
between  their local coordinates $\zeta
_i=V_i^{-1}(z)$ and $\zeta _j=V_j^{-1}(z)$, so that a complex structure 
can still be defined. The operator formalism gives a
constructive way of realizing the sewing 
procedure, and, as stressed in \cite{DPFLHS}, it generates a vertex which
embodies the geometrical structure of 
the resulting Riemann surface. It is easy to see that, in order to
generate one handle of an orientable Riemann
surface by identifying $z_i$ and $z_j$,  
a possible choice (modulo some M\"obius transformations) is
mapping the inner of the unit circle 
$C_i$ surrounding $z_i$ to the outer of the unit circle $C_j$ surrounding  $z_j
$, by  imposing: 
\begin{equation}
\label{INV}\zeta _j=\frac 1{\zeta _i} .
\end{equation}
The relation (\ref{INV}) gives the correct relation between
the orientations of the circles $C_i$ and $C_j$. 

Expanding the string field $X$ in oscillators in terms of the local
coordinates $\zeta _i$ one has:
\begin{equation}
\label{EXP}X(\zeta _i,\bar{\zeta}_i)=x-\sqrt{\frac{\alpha^{'}}{2}}p_i\ln (\zeta _i
\bar{\zeta}_i)+ i \sqrt{\frac{\alpha^{'}}{2}} \dsum\limits_{n\neq 0}
\frac 1n(\alpha
_n^{(i)} \zeta _i^{-n}+\bar{\alpha}_n^{(i)}\bar{\zeta}_i^n) .
\end{equation}
An analogous expansion in terms of $\zeta_j$ holds around $z_j$, which in
the sewing procedure must be identified 
with (\ref{EXP}) when the substitution (\ref{INV}) is made, i.e.
$$ 
\begin{array}{ccccc}
 & X(\zeta_j,\bar{\zeta}_j) & = & x-\sqrt{\frac{ \alpha^{'}}{2}}p_j\ln (\zeta _j
\bar{\zeta}_j)+i\sqrt{\frac{\alpha^{'}}{2}}\dsum\limits_{n\neq 0}\frac 1n(\alpha
_n^{(j)}\zeta _j^{-n}+\bar{\alpha}_n^{(j)} \zeta _j^n)  & \Longrightarrow \\ 
  \Longrightarrow & X'(\zeta_{i},\bar{\zeta}_{i}) & =
 & x+\sqrt{\frac{
\alpha^{'}}{2}} p_j\ln (\zeta _i \bar{\zeta}_i)+i\sqrt{\frac{\alpha^{'}}{ 2}}
\dsum\limits_{n\neq 0}\frac
1n(\alpha _n^{(j)}\zeta_i^{n}+ \bar{\alpha}_n^{(j)} \bar{\zeta}_i^{-n}). &  
\end{array}
$$
This leads to the identifications \cite{DFLS} 
\begin{equation}
\label{IDTOR}
\begin{array}{ccccc}
p_j=-p_i, &  & \alpha _n^{(j)}=-\alpha _{-n}^{(i)}=-\alpha _n^{(i)\dagger }, & 
& 
\bar{\alpha }_n^{(j)}=-\bar{\alpha}_{-n}^{(i)}=-\bar{\alpha }
_n^{(i)\dagger }.
\end{array}
\end{equation}

We defer to the original references for a more detailed description of the
remaining steps of the sewing procedure. What we want here to stress is 
that in order to construct the handle, the operators of the two Fock spaces,
corresponding to the two punctures, must be identified 
as in (\ref{IDTOR}). Successively, the trace on these oscillators is taken and
a one-loop Vertex, generating the
amplitudes on the torus, can be obtained.

In order to get instead the Klein bottle, the relative 
orientations of the two circles $C_i$
and $C_j$ must be inverted. This can be obtained 
by identifying the two punctures through the relation
\begin{equation}
\label{KINV}\zeta _j=\frac 1{\overline{\zeta _i}}.
\end{equation}
By performing the same steps as before one obtains the following identifications 
\begin{equation}
\label{IDTORO}
\begin{array}{ccccc}
p_j=-p_i, &  
& \alpha _n^{(j)}=-\bar{\alpha }_{-n}^{(i)}=-\bar{\alpha }
_n^{(i)\dagger }, &  & \bar{\alpha }_n^{(j)}=-\alpha _{-n}^{(i)}=-\alpha
_n^{(i)\dagger }.
\end{array}
\end{equation}
One then recognizes in (\ref{IDTORO}) 
(apart an irrelevant minus sign) the action of
the turnover operator $T$, which makes 
identifications between the oscillators of the holomorphic and
antiholomorphic sector. It should be not difficult, but perhaps 
tedious, to continue the sewing procedure following the same steps as described
in \cite{DPFLHS}, so to get a different 
derivation of the Klein bottle Vertex and amplitudes, given in this paper.


\begin{thebibliography}{99} 
\bibitem{CMPP}  L. Cappiello, R. Marotta, R. Pettorino and F. Pezzella, 
{\em On off-shell bosonic string amplitudes},
hep-th/9804032, accepted for publication in {\em Mod. Phys. Lett.} {\bf A}. 

\bibitem{BK}  Z. Bern and D. A. Kosower, {\em Phys. Rev.} {\bf D38}
(1988) 1888; {\em Nucl. Phys.} {\bf B321} (1989) 605; {\em Nucl. Phys.} 
{\bf B379} (1992) 451.

\bibitem{DMLRM}  P. Di Vecchia, R. Marotta, A. Lerda, R. Russo and L. Magnea, 
{\em Nucl. Phys.} {\bf B469} (1996) 235.

\bibitem{BDS}  Z. Bern, D. C. Dunbar and T. Shimada, {\em Phys. Lett.} 
{\bf B312} (1983) 277.


\bibitem{S}  J.H. Schwarz, {\em Phys. Rep.} {\bf 89} (1982) 223.

\bibitem{HV} G. 't Hooft and M. Veltman, {\em Ann. Inst. Henri Poincar\`{e}},
vol. XX, n. 1, 1974, 69-94.

\bibitem{DPFLHS}  P. Di Vecchia, F. Pezzella, M. Frau, A. Lerda, K.
Hornfeck and S. Sciuto, {\em Nucl. Phys.} {\bf B322} (1989) 317.

\bibitem{HKKS}  M. Hayashi, N. Kawamoto, T. Kuramoto and K. Shigemoto, {\em Nucl. Phys.} 
{\bf B296} (1988) 373;

J. Liu, {\em Nucl. Phys.} {\bf B362} (1991) 141;

V. A. Kostelecky, O. Lechtenfeld and S. Samuel, {\em Nucl. Phys.} {\bf B298}
(1988) 133.

\bibitem{M}  E. Martinec, {\em Nucl. Phys.} {\bf B281} (1987) 157.

\bibitem{RS}  K. Roland and H. Sato, {\em Nucl. Phys.} {\bf B480} (1996) 99.

\bibitem{GSW}  M. Green, J. Schwarz, E. Witten, {\em Superstring Theory},
Cambridge University Press (1987).

\bibitem{BM}  C.P. Burgess and T.R. Morris, {\em Nucl. Phys.} {\bf B291}
(1987) 285.

\bibitem{DFLS} P. Di Vecchia, M. Frau, A. Lerda and S. Sciuto, {\em Phys.
Lett.} {\bf B199} (1987) 49.
\end{thebibliography}
\end{document}